# Spectral Efficiency Considerations for 6G

Joseph Boccuzzi, *Member, IEEE*

***Abstract*— As wireless connectivity continues to evolve towards 6G, there is an ever-increasing demand to not only deliver higher throughput, lower latency, and improved reliability, but also do so as efficiently as possible. To this point, the term efficiency has been quantified through applications to Spectral Efficiency (SE) and Energy Efficiency (EE). In this paper we introduce a new system metric called Radio Resource Utilization Efficiency (RUE). This metric quantifies the efficiency of the available radio resources (Spectrum, Access Method, Time Slots, Data Symbols, etc.) used to deliver future 6G demands.**

**We compare the system performance of Typical Cellular and Cell-Free Massive MIMO deployments as a vehicle to demonstrate the need for this new metric. We begin by providing a concise treatment of items impacting SE by introducing three categories: 5G Radio Resources, Practical Limitations (such as channel matrix rank deficiency) and Implementation Losses (SINR degradation).**

**For the example Radio Access Technology configuration analyzed, we show 5G yields an RUE of 47% (revealing significant room for improvement when defining 6G). Practical limitation assumptions are compared to 5G Multi-User MIMO (MU-MIMO) measurements conducted in a commercialized deployment. SE losses are characterized to offer guidance to advanced algorithms employing Machine Learning (ML) based techniques.**

**We present the benefits of increasing the transmission Bandwidth (BW) from 100MHz to 1.6GHz. We describe a Next Generation RAN architecture that can support 6G and AI-RAN.**

## I. INTRODUCTION

We have embarked into an Era where the intersection of certain technologies need to be considered when defining the next generation Telecom network. Excitedly, these concerned technologies are: 5G/6G evolution, Generative Artificial Intelligence (AI) and High-Performance Computing (HPC) & Networking. These technologies collectively represent a paradigm shift in how we will communicate in the future.

The evolution of 5G is addressed on various fronts. We begin with the traditional roadmap items such as higher throughput, lower latency, increase user density, improve energy efficiency and increase spectral efficiency. As we prepare for 6G we will introduce new technologies such as Cell-Free Massive MIMO [1], [2], [3], [4] and Reconfigurable Intelligent Surfaces (RIS) [5], [6], both of which improve user performance at the cell edge. To achieve higher throughputs and an increase in spectral efficiency we will need to: increase Signal-to-Interference-plus-Noise Ratio (SINR), utilize wider transmission BW (at higher frequency bands), increase the number of MIMO layers (expand massive MIMO capabilities) and use Carrier Aggregation.

Next, we see AI continues to bring forth new opportunities while significantly impacting operational and management efficiency. Generative AI is exponentially increasing in terms of usage as well as capability/complexity. We have seen ChatGPT reach 100 million users in ~2 months, easily outpacing other popular applications such as TikTok. The capabilities of the Generative AI models have exponentially increased in a past few years from the introduction of the transformer to increasing the number of AI foundational model parameters from hundreds of billions to trillions of parameters. The availability of HPC & Networking enables SW-defined solutions that ignite researchers' abilities to innovate at an exponential rate.

Lastly, the adventures in HPC & Networking, from both the Software (SW) & Hardware (HW) perspectives, are expected to quickly become adopted in the Telecom industry. Larger Cloud Service Providers (Hyperscalers) were essentially caught off guard to the quick adoption of these generative AI models. As the AI foundational model parameter size increases, more computational processing capability is needed to reasonably train and perform inference. This computational processing capability can take on the form of an AI Factory which consists of a large amount of Graphics Processing Units (GPUs) that need efficient networking to support the increase in traffic.[7] It is important to note networking should not introduce latency especially when collecting/transferring data between different stages of the foundational model. Low latency can be accomplished by a network that delivers packets in sequence order, predicts and adapts to traffic congestion. These networking capabilities are supported in the NVIDIA Spectrum X platform.[8], [9]

The next generation Telecom network architecture needs to consider these multi-dimensional technology requirements during its definition. The next generation Telecom network needs to be flexible, scalable, efficient and cloud native – all pointing to a SW-defined, End-to-End (E2E) network.

We expect the continued success of Open and Standardized interfaces to help proliferate 6G deployments. Standardization groups such as 3GPP [10] & O-RAN Alliance [11] are essential to the ecosystem to enable collaboration in designing the best product. Moreover, this helps accelerate innovation in

This paragraph of the first footnote will contain the date on which you submitted your paper for review, which is populated by IEEE. It is IEEE style to display support information, including sponsor and financial support acknowledgment, here and not in an acknowledgment section at the end of the article.

Joseph Boccuzzi is with NVIDIA Corp., Santa Clara, CA 95052 USA (e-mail: jboccuzzi@nvidia.com).

Color versions of one or more of the figures in this article are available online at http://ieeexplore.ieee.org

technology needed to deliver the future use cases & Key Performance Indicators (KPIs). We feel it is only with open interfaces and SW-defined networks that 6G can benefit from the same exponential growth seen in the Generative AI field. It is for these reasons that the AI-RAN Alliance has been created to bring together technology leaders to enhance the performance and capability of the RAN with AI.[12]

Figure 1 shows a 5G network based on 3GPP & O-RAN standards. We placed the user applications (denoted as Generative-AI based) in a data center cloud connected to a 5G stand alone Core Network (CN). The 5G gNB consists of O-RAN Distributed Units (O-DU) and O-RAN Centralized Units (O-CU). The 5G gNB interfaces to the Service Management & Orchestration (SMO) via the interface labeled O1 and RAN Intelligent Controller (RIC) via the E2 interface. The SMO is an intelligent RAN automation platform to address life cycle management, network performance, etc. The RIC is responsible for controlling and optimizing RAN functions such as improving QoS, traffic steering, user capacity, etc. The RIC is divided into non real-time (Non RT-RIC) and near real-time (Near RT-RIC) components. O-RAN defines the Front Haul (FH) to consist of Control, User, Synchronization & Management – Planes (C/U/S/M-Planes). The O-RAN Radio Unit (O-RU) converts the digital FH to radio signals. Accurate time synchronization is essential in a communication system, a Grand Master serves as the primary time source to ensure accurate time synchronization across connected systems.

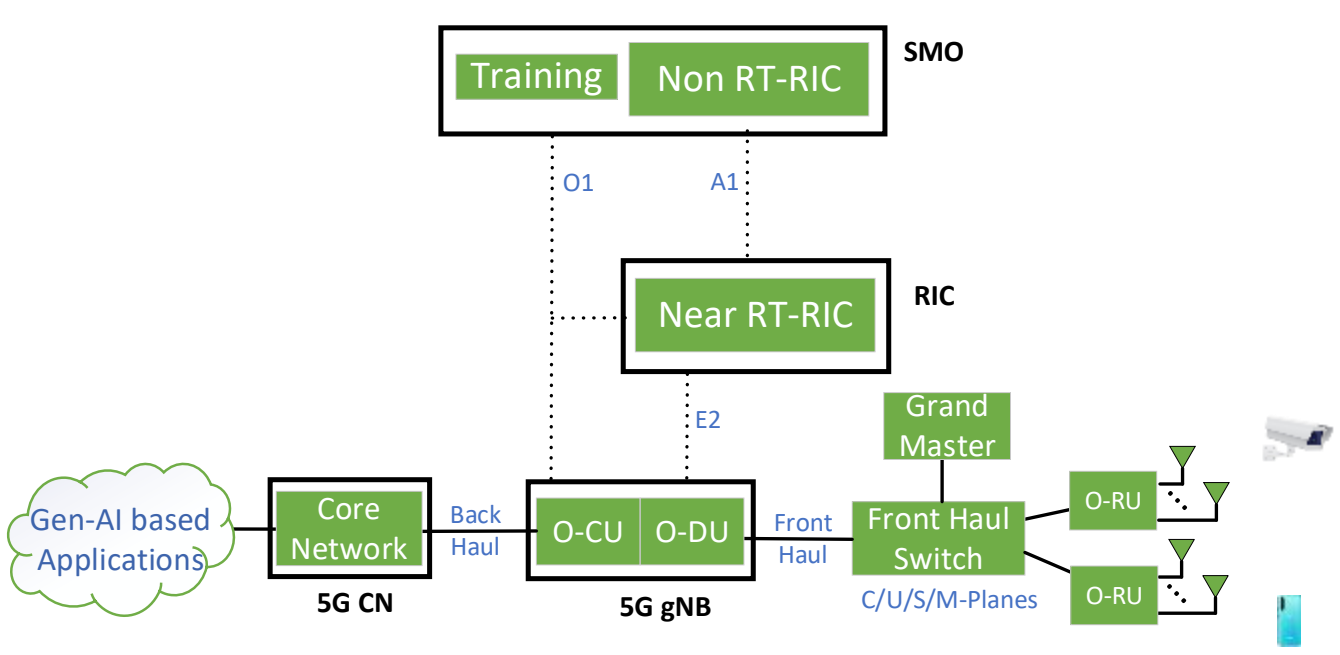

**Fig. 1.** Example 5G network highlighting 3GPP and O-RAN defined interfaces.

We will present two example deployment scenarios and provide expected performance improvements with this next generation Telecom network. In a traditional cellular deployment, the coverage area is divided into cells to deliver signals to the desired users. Adjacent cells using the same carrier frequency will experience interference from neighboring cells and introduce interference to neighboring cells. This interference is predominantly seen at the cell edge which will exhibit the lowest SINR values. Consequently, this will impact user throughput at the cell edge. In other words, users near the center of the cell will experience higher throughput while users at the edge of the cell will experience lower throughput. In fact, the SINR statistics are widely varying in between thus delivering non-uniform QoS throughout the coverage area. If we assume a uniform distribution of users within a cell, the largest population will be on a circular band at the edge. Deploying Cell-Free Massive MIMO and RIS technologies will not only deliver improved performance at the cell edge, but also a more uniform QoS throughput the coverage area.

Deploying MIMO improves the SINR allowing for higher throughput and improved coverage. Consider using MIMO for spatial multiplexing where the user throughput will increase by transmitting parallel data streams concurrently. However, the wireless channel matrix rank needs to be supportive. The wireless channel is random and can experience occasions when it will not support full rank or peak throughput communication. In later sections, field trial results will be reviewed to support this discussion. For example, consider a 4x4 MIMO deployment that experiences full rank (rank = 4) about 40% of the time, indicating 60% of the time 4x4 spatial multiplexing MIMO is not achieving its maximum potential (we use the notation 60/40% for this example).

In this paper we will utilize an example 5G network configuration to identify obstacles that limit advancement in Spectral Efficiency (SE). Their performance impact is evaluated, and improvements are suggested for 6G. In doing so, we will identify which obstacles cause the largest degradation that should be addressed in the definition of 6G. To help introduce and evaluate these obstacles we will only consider the downlink (DL) direction.

It is important to understand: "What is impacting SE in 5G?" The answer is provided with a concise summary by introducing three categories: 5G Radio Resources, Wireless Channel Rank Statistics and Implementation Loss.

- 5G Radio Resources - considers how efficiently we utilize the resources to deliver DL traffic. For example, this category includes the Time Division Duplex (TDD) slot format, number of data symbols used in the time slots, number of physical resource blocks available, etc.
- Wireless Channel Rank Statistics - considers the occasions when full rank transmission is not supported by the wireless channel.
- Implementation Loss - considers degradations from channel estimation, frequency offsets, Forward Error Correction (FEC) decoder, etc.

We begin by providing a review of theoretical SISO based SE results for two deployment scenarios: Typical Cellular and Cell-Free. These results are presented here to highlight assumptions made, for ease of comparison and to establish the performance baseline which we will expand upon as our discussions progress.

Next, we extend the SISO based SE results to support 4x4 MIMO and apply the respective limitations & losses to these two deployment scenarios. These results reveal the largest impact to SE performance is from the methods 5G delivers services. Hence, we have identified one area that requires more focused attention which is the ability of the radio resources to deliver services efficiently. We introduce a new system metric called Radio Resource Utilization Efficiency (RUE) to quantify this inefficiency so it can be improved in 6G. A proposed RUE target value for 6G is given. We provide suggestions in each of the three categories to overcome the respective items impacting SE. Lastly, we provide an AI-RAN network architecture that is configured, as an example, for a cell-Free deployment.

The contributions of this paper:

- ⇨ Introduced a concise treatment of three categories that impact SE in 5G deployments.
- ⇨ Provided a review of SISO/MIMO theoretical SE performance for two example deployment scenarios.
- ⇨ Evaluated the impact of SE due to the practical limitations and implementation losses present in 5G.
- ⇨ Introduced the new system metric called RUE.
- ⇨ Discussed impact of larger transmission BWs on SE and peak cell throughput.
- ⇨ Compared our theoretical results to real-world measurements from commercial 5G deployments.
- ⇨ Identified areas of improvement and provided suggestions to be addressed in the definition of 6G cellular technology.
- ⇨ Proposed a Network Architecture to support the co-existence of Typical Cellular and Cell-Free deployments with ability to grow with 6G and support AI-RAN.

The remainder of this paper is organized as follows. Section II details the analysis and assumptions used in modeling both typical cellular and Cell-Free deployments. Section III extends this analysis to incorporate MIMO systems. Section IV presents 3 categorizes impacting SE. Section V introduces the novel RUE metric. Section VI quantifies the SE impact of these 3 categories. Section VII examines the performance gains of increasing system BW and introduces measurements from commercial 4G & 5G networks. Section VIII presents a network deployment architecture to support AI-RAN. Finally, Section IX summarizes the key conclusions of this work.

## II. Deployment Analysis & Assumptions

In this section we provide analytical SE results for SISO applications. We begin by describing the spatial cellular layout of L=9 cells, shown in Figure 2. Each cell is denoted by a black dot and equipped with an omni directional antenna. The cells are separated in distance denoted by Inter Site Distance (ISD), set to 200 meters. Figure 2a shows the center cell as the desired cell and the surrounding 8 cells are contributing interference. Here a user located in the desired area is only being serviced by the center cell. The remaining "L-1" cells act as interference to the desired user. Figure 2b shows all the cells are contributing to the desired signal power.

All cells in Figure 2 transmit with the same power, cells in 2a transmit with power, $P_t$, whereas cells in 2b transmit with power $P_t/9$. This constraint is used to support a fair comparison between the two deployment options. All the cells simulated in this paper will experience downlink path loss, modelled according to the Close-in (CI) path loss model defined as: [13]

$$PL(d) = FSPL(d_0 = 1m) + 10 \cdot n \cdot Log\left(\frac{d}{d_0}\right) \quad (1)$$

Where: FPSL = Free Space Path Loss, n = Path Loss Exponent (n = 3.52) and PL(d) = Path Loss at distance d meters expressed in dB. We have assumed no user is within a radius of 5m of the O-RU. The impact of noise (N) is introduced as follows:

$$N\ (dBm) = -173.6 + 10 * \mathrm{Log}(BW) + NF \quad (2)$$

Considering a BW = 100MHz and Noise Figure (NF) = 5dB, (2) arrives with a Noise Power of -88.6dBm which is used in the simulations. The desired user occupies a BW of 100MHz.

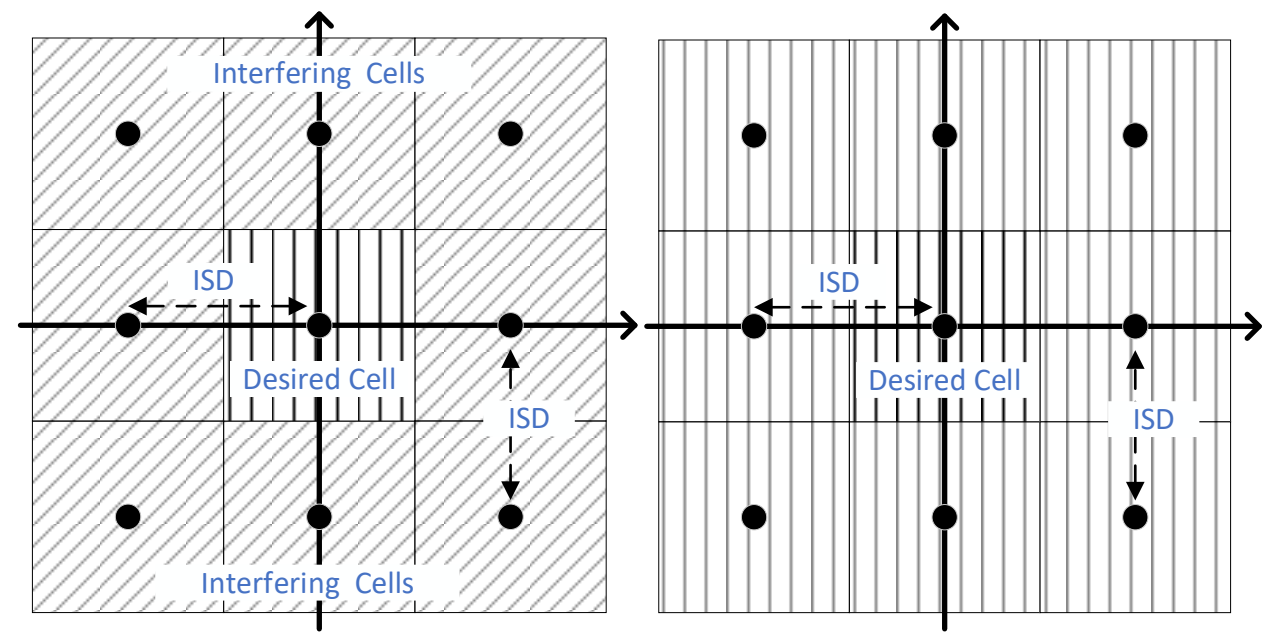

**Fig. 2.** Example 9 Cell layout: (a) Left side represents Typical Cellular (b) Right side represents Cell-Free.

We evaluate two deployments scenarios in this analysis. These results are presented here for ease of comparison and to establish the performance baseline which we will expand upon as our discussions progress. The first scenario models the Typical Cellular deployment where neighboring cells act as interference to the center cell, in this case we need to consider SINR. This deployment is shown in Figure 2a. The second scenario models a Cell-Free deployment where all cells are part of the desired signal, in this case we need to consider Signal-to-Noise Ratio (SNR). This deployment is shown in Figure 2b.

The major reason for making these assumptions in this section is so we can focus on the main concepts and evaluate the theoretical performance instead of including practical constraints (which will be addressed in the later sections).

### A. *Typical Cellular Deployment*

In a typical cellular deployment, all the antennas used to communicate to the users in a targeted area are included as part of a single O-RU. In this section each user and O-RU will have a single antenna, this is the SISO use case. We will derive the SINR of a single user located in the desired cell. The downlink (DL) signal received at the $k^{th}$ UE (within the desired cell corresponding to l=1) is written as:

$$y_k = h_{1k} \cdot s_{1k} + \sum_{l=2}^{L} h_{lk} \cdot s_{lk} + n_k \quad (3)$$

Here $h_{lk}$ denotes the channel from $l^{th}$ O-RU to the $k^{th}$ user and $n_k$ is the receiver noise of the $k^{th}$ user. The total number of O-RUs included in the analysis is L, where L-1 are considered interference cells. The transmitted signal from the L-1 O-RUs to the $k^{th}$ user is $s_{lk}$. The power of the desired signal (l=1) is:

$$E\{|s_{1k}|^2\} = \sigma_s^2 \quad (4)$$

Applying a weight, $w_{1k}$, on the received signal of the k[th] UE within the desired cell, we define a new variable, $x_k$ (* denotes the complex conjugate operation) $x_k = w_{1k}^* \cdot y_k$:

$$x_k = w_{1k}^* \cdot h_{1k} \cdot s_{1k} + \sum_{l=2}^{L} w_{1k}^* \cdot h_{lk} \cdot s_{lk} + w_{1k}^* \cdot n_k \qquad (5)$$

With the receiver weight defined as $w_{1k} = \frac{h_{1k}}{|h_{1k}|}$, the DL SINR observed by the k[th] UE in the desired cell (l=1) is represented as:

$$SINR_k = \frac{\sigma_s^2 \cdot |h_{1k}|^2}{\sigma_l^2 \cdot \sum_{l=2}^{L} |h_{lk}|^2 + \sigma_n^2} \qquad (6)$$

Where $\sigma_n^2$ is the noise power. The channel capacity, C, can then be represented as:

$$C = Log_2 \left[ 1 + \frac{\sigma_s^2 \cdot |h_{1k}|^2}{\sigma_l^2 \cdot \sum_{l=2}^{L} |h_{lk}|^2 + \sigma_n^2} \right] \qquad (7)$$

Channel capacity, C, is also referred to Spectral Efficiency (SE) with the units of bps/Hz/Cell.

B. *Cell-Free Deployment*

In this user centric, Cell-Free deployment, all the antennas used to communicate to the users in a targeted area are not located in a single O-RU.[14], [15], [16] They are in fact distributed across many O-RUs. In this section we consider using multiple O-RUs to transmit the desired signal to a single user. We will derive the SNR of a single user located in the center of the desired coverage area. Since we want the signals from all the L gNBs to collectively present the desired signal, we apply a precoder weight, $w_{lk}$, on the downlink (DL) transmit signal and define a new variable, $x_k$. Here $s_{lk}$ represents the desired signal from the l[th] O-RU to the k[th] user.

$$x_{lk} = w_{lk}^* \cdot s_{lk} \qquad (8)$$

The downlink signal received at the k[th] UE, considering the contribution of all the L O-RUs is written as:

$$y_k = \sum_{l=1}^{L} h_{lk} \cdot x_{lk} + n_k \qquad (9)$$

$$y_k = \sum_{l=1}^{L} h_{lk} \cdot w_{lk}^* \cdot s_{lk} + n_k \qquad (10)$$

The number of O-RUs involved in serving the UE is equal to L, where are all the O-RUs transmit the desirable signal. The power of all the transmitted Cell-Free signals, $s_{lk}$, is normalized to the power of a single Typical Cellular O-RU, $\sigma_s^2$.

$$E\{|s_{lk}|^2\} = \sigma_{sl}^2 = \frac{\sigma_s^2}{L} \qquad (11)$$

With the transmitter precoder weight defined as $w_{lk} = \frac{h_{lk}^*}{|h_{lk}|}$, the DL SNR observed by the k[th] UE at the center of the coverage area is provided by:

$$SNR_k = \frac{1}{\sigma_n^2} \cdot \sum_{l=1}^{L} |h_{lk}|^2 \cdot \sigma_{sl}^2 \qquad (12)$$

The interference from cells located outside the L contributing cells is not considered, as the expected propagation path loss is very significant. The channel capacity, C, is represented as:

$$C = Log_2 \left[ 1 + \frac{1}{\sigma_n^2} \cdot \sum_{l=1}^{L} |h_{lk}|^2 \cdot \sigma_{sl}^2 \right] \qquad (13)$$

C. *Deployment Comparison: Typical Cellular vs. Cell-Free*

Next we compare the DL SISO system performance between the two deployments in Figure 3. Here the Cell-Free (dashed line) exhibits better performance over typical cellular (solid line). Better performance has 2 characteristics: first the curve should shift to the right indicating larger SINR/SNR values are encountered. Second, the curves should be steeper in slope indicating less variation of SINR/SNR values encountered. This second characteristic leads to a more uniform QoS throughput the coverage area.[2] The curves on the right side show the respective DL SINR/SNR values, expressed in dB, within the cell. The curves on the left side show the corresponding SE expressed in bps/Hz.

When considering the SINR CDF plot on the right side, we make the following observations on the maximum and minimum values. The maximum SINR value occurs at the center of the cell, where the adjacent cell interference gets significantly decreased due to the increase in propagation path loss. The smaller SINR values occur at the edge of the cell, where the interference is significant due to the proximity of the adjacent cells.

To present the performance gains we consider two points on the CDF (highlighted by circles on the graph): 10% and 90%. Consider the Cell-Free SE plot on the left side where CDF = 0.9, this means 90% of the values are less than 6.83 bps/Hz. From a Complimentary CDF (CCDF) perspective, this implies that 10% of the SE values are greater than 6.83 bps/Hz. Similarly, CDF = 0.1 indicates 10% of the SE values are less than 2.47 bps/Hz. From a CCDF perspective, this implies that 90% of the SE values are greater than 2.47 bps/Hz.

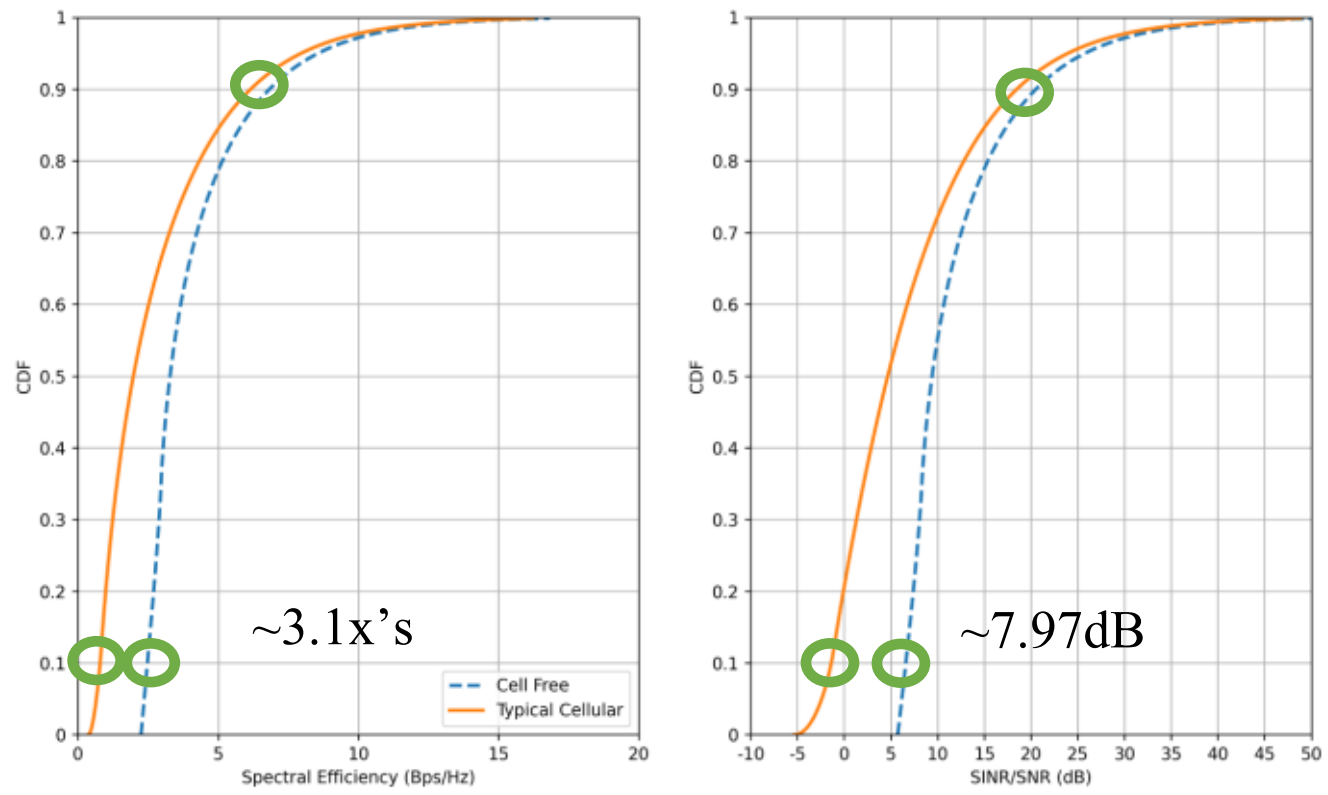

**Fig. 3.** SE & SINR/SNR performance comparison of Conventional and Cell-Free deployments.

The SE & SNR performance gains of Cell-Free over Typical Cellular are summarized in Table 1. The results show approximately 3.1x's gain in SE can be achieved with Cell-Free in greater than 90% of the cell (when compared to Typical Cellular). The SE gain is more at this operating point (compared to 10%), since it is associated with the cell edge (due to the lower SINR/SNR values).

TABLE I
Cell-Free Gains over Typical Cellular deployment.

| CCDF | SE Gain | SNR Gain |
|---|---|---|
| 10% | 6.83/6.16 = 1.1x's | 20.53 – 18.48 = 2.05dB |
| 90% | 2.47/0.79 = 3.1x's | 6.6 – (-1.37) = 7.97dB |

## III. 4x4 MIMO Performance

We have provided results for SISO and now would like to discuss 4x4 MIMO. We increase the number of antennas at the O-RU and the k$^{th}$ user to 4. The intension is using MIMO to increase the DL throughput. This is accomplished by extending the above SISO results to support up to M spatial layers by using an upper bound for SE given below:

$$C = Log_2\left[det\left|I + \frac{SNR}{M} \cdot H \cdot H^*\right|\right] \quad (14)$$

Which can be rewritten using the Eigen Spectral decomposition of the Channel Matrix, H.

$$C = \sum_{i=1}^{M} Log_2\left[1 + \frac{SNR}{M} \cdot \lambda_i^2\right] \quad (15)$$

Forcing all eigen values to equal a constant ($\lambda_i^2 = M$) creates a full rank scenario where M spatial channels can be supported. With these assumptions, we arrive with the upper bound channel capacity, C, given as follows:

$$C \leq M \cdot Log_2[1 + SNR] \quad (16)$$

It would be very optimistic to assume the channel matrix, H, enables a full rank 100% of the time, in other words assuming Probability (rank 4) = 100% (over the entire cell).[17], [18] Several factors can influence the number of layers a system can support, for example antenna correlation, poor scattering environment, etc. We provide an example channel matrix rank distribution by assuming Probability (rank 2) = 60% and the Probability (rank 4) = 40%. This channel matrix rank assumption is reasonably close to the measurements made in a 5G Massive MIMO commercial deployments.[17], [18]

Using the upper bound channel capacity and the assumed channel matrix rank distribution, we compare SE in Figure 4. The right side shows the SE for the Typical Cellular case, while the left side shows the same comparison for the Cell-Free case. For both deployment cases we consider SISO, 4x4 MIMO using 60%/40% rank assumption and the ideal 4x4 MIMO.

The impact of using MIMO is more visible (when comparing the distance between the 3 curves) in Cell-Free than in Typical Cellular. However, the spatial multiplexing gain is the same for both deployments as provided in (16).

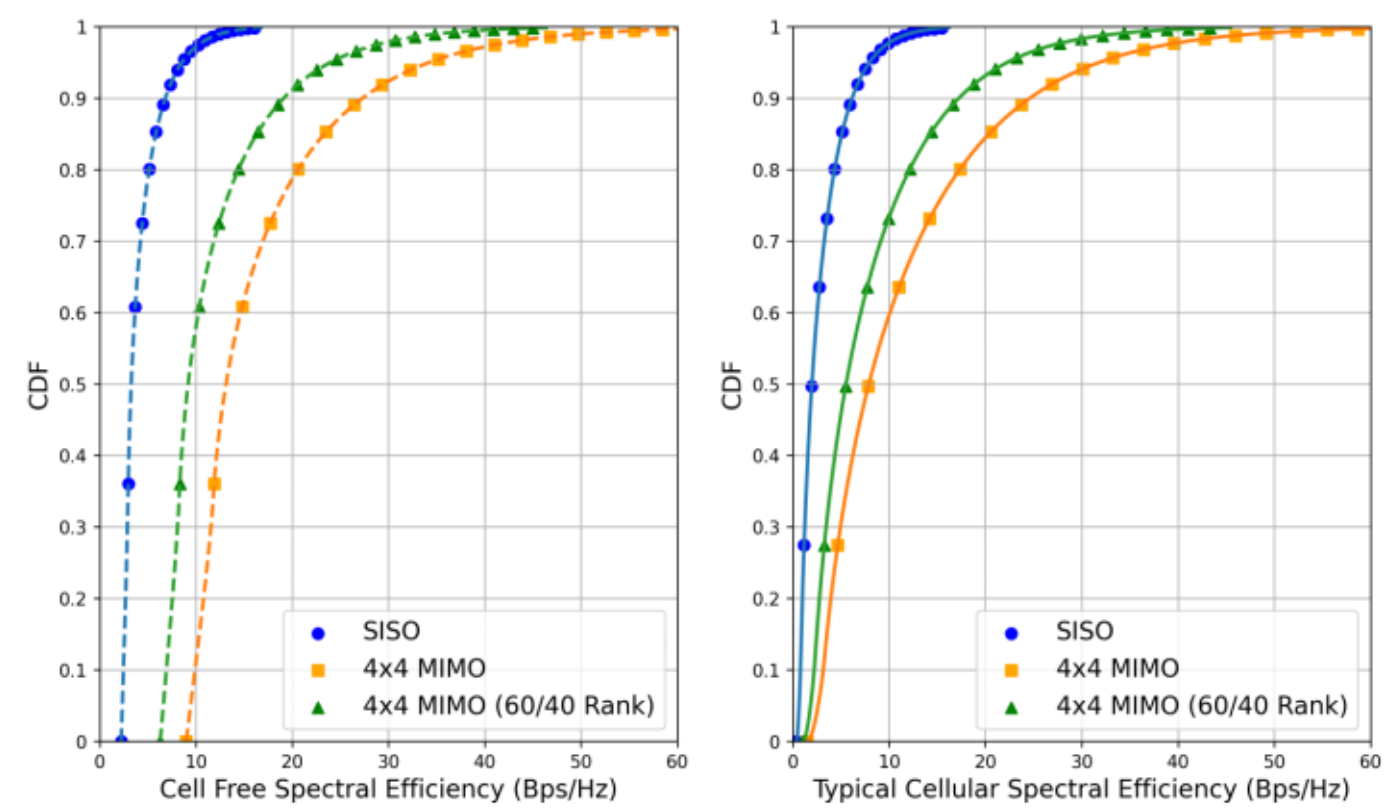

**Fig. 4.** SISO & MIMO SE Comparisons with assumed Channel Matrix rank distribution.

Using the upper bound for spatial multiplexing assumption, we applied a gain of M = 4 to the SISO SE results for the 4x4 MIMO scenario. When using the 60/40% assumed matrix rank distribution, the gain reduces from 4x's to approximately 2.8x's the SISO SE results. Using MIMO significantly improves SE, even with the assumed channel matrix rank deficiency.

## IV. Practical Limitations and Implementation Losses

### A. *Items for Considerations*

In this section we will consider practical limitations and implementation losses impacting SE performance. Specifically, we will discuss 3 categories: Implementation Loss, Wireless Channel Rank deficiency and 5G Radio Resources.

**Implementation Loss (IL):**

We define IL to be the dB difference (of SNR/SINR) between theoretical and measured performance. IL is specific to vendor implementations and often considered to be a component of product differentiation. These losses include RF design impairments (non-linearities, group delay, etc.), quantization effects, channel estimation inaccuracy, presence of timing and frequency offsets, measurement errors, # of FEC decoder iterations, MAC scheduler efficiency, etc.[19] These losses contribute to the degradation in SNR/SINR which leads to a degradation in SE. We will consider this loss to be uniformly distributed across the SNR/SINR performance. We model the implementation loss by intentionally introducing a constant offset in dB (from 3dB to 0.5dB) to the ideal performance curves. This allows us to address multi-vendor losses in a general manner yet provide direction on potential improvements. In this study we consider 3dB to be the worse degradation. We vary the IL to identify the potential SE gains using advanced signal processing algorithms.

**Channel Matrix Rank Deficiency:**

These losses are due to the channel matrix, H, having less than full rank, in this case rank = 4 (R = 4) for 4x4 MIMO. We provided an example rank distribution (60/40) which significantly reduced the full rank gain from 4x's to that of 2.8x's, compared to SISO. This value was obtained by $P(R=2) \cdot SE(R=2) + P(R=4) \cdot SE(R=4)$, where P(R = 2) represents the Probability the rank = 2.

We assumed each cell to have the same rank deficiency behavior. Recent results on a commercial 5G network in Dallas, Texas reveal the average SU-MIMO layers measured to be approximately 3.4 in the downlink.[17]

Separately, other Real-World measurements were made on three commercial 5G (& 4G) networks on low and mid frequency bands. These results showed the channel matrix rank rarely supports the full 4 layers of a 4x4 MIMO link.[18] We averaged 6 data sets, and the results revealed the probability of having a full rank was approximately 7% and the overall channel rank was approximately 2.5.

**5G Radio Resources Practical Limitations:**

These losses include realistic system configurations used in practical deployments that degrade SE, such as:

- **# of active DL slots**
  - Highly dependent on the TDD frame structure chosen.
  - For an example frame structure of DDDSUUDDDD (where D=DL, U=UL & S=Special Slot), we assume 28 out of 40 slots are used for DL, which results in **70%** (=28/40) usage of time slot resources.
  - We have chosen the DL-Heavy slot format since the focus of this work is on the DL only.
  - If we consider a frame structure of DDDSU, then our usage reduces to **60%** (=24/40).
- **# of actual data symbols**
  - Of the 14 symbols available in a slot, we will assume 11 data symbols are available for Physical DL Shared Channel (PDSCH) + 1 Pilot + 2 Physical DL Control Channel (PDCCH), which results in **78.6%** (=11/14) usage of downlink symbol resources.
  - We consider 1 Pilot symbol to be the minimum deployment option.
  - If we consider 3 Pilot symbols are used, then our usage reduced to **64%** (=9/14).
- **FEC Code Rate**
  - Peak throughput is achieved with the largest MCS index supporting 256QAM, this results in **0.92578** (=948/1024).[20], [21]
  - Please note for the lower SNR values, lower code rates are used thus lowering the user throughput even further.
- **# of Physical Resource Blocks (PRBs)**
  - Instead of considering the entire 100MHz, assume **98.2MHz** utilization (=273 PRBs * 30KHz SCS * 12 RE/RB). Where: SCS = Sub-carrier Spacing, RE = Resource Element & RB = Resource Block.
  - Here we used the maximum allowable PRBs.
- **Cyclic Prefix (CP) Overhead**
  - A time slot consists of 14 symbols, but when considering the overhead of inserting the CP, the overall slot duration is equivalent to 15 Symbols. This results in 14/15 or **93.3%** of actual usage.

When including these practical limitations, they will lead to approximately 50% reduction in downlink user throughput (47% = 0.92578 * 0.7 * 0.7857 * 0.982 * 0.933). If we consider the DDDSU frame structure and 3 Pilot symbols per slot, this value drops to ~33% (= 0.92578 * 0.6 * 0.64 * 0.982 * 0.933). This means the 5G radio resources have significantly reduced the achievable throughput before any channel impairment degradation and implementation losses are considered.

### B. *Including Practical Limitations & Implementation Losses*

Based on the previous section, a realistic scenario would consist of applying a 50% reduction, due to practical limitations, include a 3dB implementation loss (assumed worse case) and utilize the assumed 60/40 channel matrix rank distribution. Figure 5 shows the SE performance of 4x4 MIMO including practical and implementation losses. The curves on the right side correspond to Typical Cellular, whereas the curves on the left side correspond to Cell-Free. The results show the incremental impact of including 50% reduction due to practical 5G radio resource limitations, followed by the assumed 60/40 matrix rank distribution and lastly, adding a 3dB IL. The biggest loss was observed by considering the practical 5G radio resource limitations. The second largest loss was the wireless channel rank deficiency, here assumed to be 60/40. Lastly, the IL contributed the smallest loss of the 3 categories considered.

Focusing on the CDF = 10% operating point we can make 2 observations: First, without any IL, the Cell-Free SE gain remains at 3.1x's that of Typical Cellular deployments. This is because the MIMO gain, radio resource loss and 60/40 channel matrix rank deficiency have their contributions modeled by a linear multiplier. Second, when considering IL = 3dB, the Cell-Free SE gain is now 3.8x's that of Typical Cellular. This is expected since the slope of the SE curve is steeper in this region.

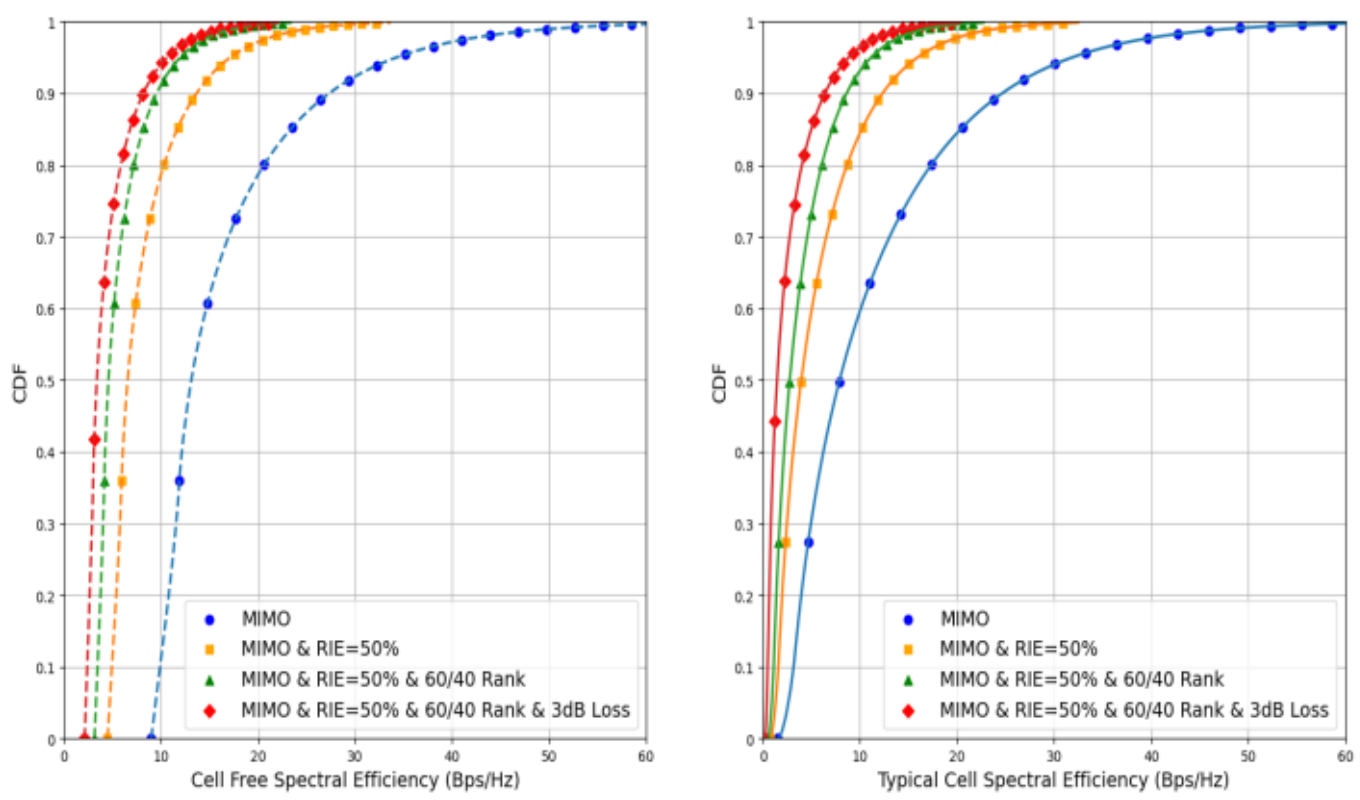

**Fig. 5.** 4x4 MIMO SE performance considering practical and implementation losses.

To present the incremental losses due to all the above-mentioned practical items, we provide an alternative view in Figure 6. Here we plot SE vs. SNR for each of these scenarios:

- Full Rank 100% (Ideal).
- 50% reduction due to 5G radio resources.
- Assumed 60/40 channel rank distribution.
- 3dB implementation loss.

We can clearly see 5G radio resources provide the largest SE loss of 50%. Next using the assumed rank distribution adds an incremental 15% of SE loss, totaling 65%. Lastly, adding IL = 3dB introduces an incremental loss of 5% at the higher SNR range (indicative of the cell center) and incremental loss of 15% at the lower SNR range (indicative of the cell edge). Recall the earlier discussion on slope of SE curve is steeper in this lower SNR range.

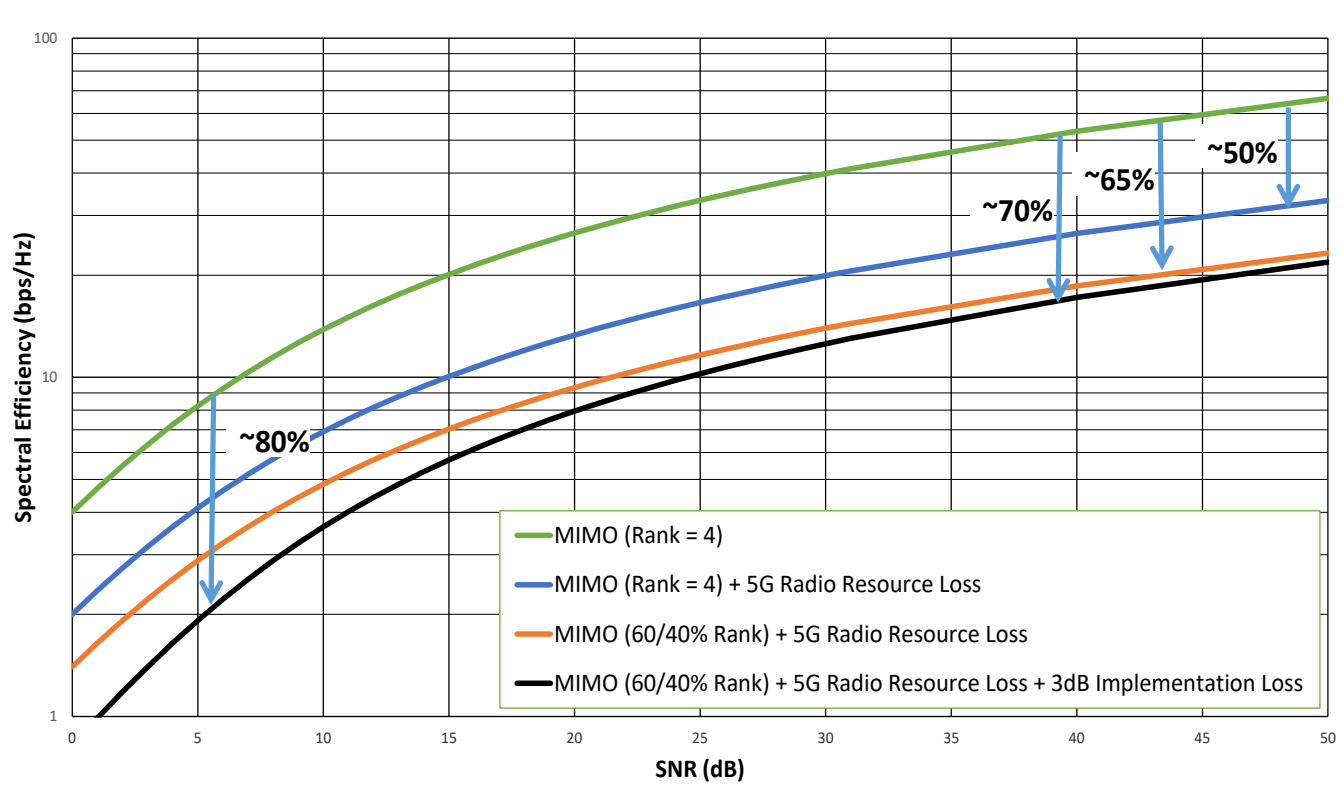

**Fig. 6.** MIMO Spectral Efficiency vs. SNR performance comparison.

We discuss the potential SE gains possible with improvements to receiver signal processing algorithms.[22], [23] These improvements are a result of introducing ML-enhanced processing such as applying a neural network to channel estimation, equalization, symbol de-mapping, etc. The potential SE gains are considered by starting with a 3dB loss in SNR/SINR and then reducing that loss to 2dB, 1dB and finally to 0.5dB. The SE gains are relative to the 3dB implementation loss system performance. The potential SE performance improvements, achieved by a reduction in implementation loss, is provided in Table 2. The results correspond to greater than 90% of the area and show that solutions with the capability to provide receiver signal processing improvements can result in a SE gain of 36% for Cell-Free and 61% for Typical cellular.

Table 2 reveals the tremendous potential with ML-enhanced solutions capable to reduce the implementation losses. However, please recall this is the smallest overall contribution when considering 5G radio resources and channel rank deficiency, as shown in Figures 5 and 6.

TABLE II

SE Gains due to improvements in Implementation Loss.

| Implementation Loss Improvements | Typical Cellular | Cell-Free |
|---|---|---|
| Delta = 1dB (from 3dB to 2dB) | 22% | 14% |
| Delta = 2dB (from 3dB to 1dB) | 47% | 29% |
| Delta = 2.5dB (from 3dB to 0.5dB) | 61% | 36% |

In other words, there was about an overall 5% loss from "MIMO (60/40) + 5G radio resources" to "MIMO (60/40) + 5G radio resources + 3dB IL" at high SNR and about overall 15% for the low SNR values. For example, 2.5dB improvement in IL delivers a gain of 36% (compared to 3dB). This will translate to an overall gain of (36%) * (15%) = ~5.4% for the low SNR values and (36%) * (5%) = ~2% for high SNRs.

## V. Radio Resource Utilization Efficiency (RUE)

It is for the above reasons we have introduced the new system metric, Radio Resource Utilization Efficiency (RUE) to help increase the focus and attention the wireless system deserves. This metric is unitless and provides an indication of how efficiently all the defined resources are being used to provide the desired services. We have defined radio resources to include Spectrum, Access Method, FEC Code Rate, Time Slots, Data Symbols, etc.

$$RUE = \frac{Allocated\ Resources}{Total\ Number\ of\ Resources} \quad (17)$$

This implies radio resource loss (or inefficiency) to be defined as 1 – RUE. For the example 5G scenario, the RUE = 47% (= 0.92578 * 0.7 * 0.7857 * 0.982 * 0.933) which is calculated as:

$$RUE = \binom{Spectrum}{Used} \cdot \binom{Time\ Slots}{Allocated} \cdot \binom{Symbols\ Used}{for\ Data} \cdot \binom{FEC}{Code\ Rate} \cdot \binom{Samples\ that}{contain\ Data} \quad (18)$$

We are suggesting this new Efficiency metric, RUE, be considered in the radio resource definition of 6G cellular technology. Three categories that impact SE have been presented: Radio Resource Efficiency, Channel Matrix Rank and Implementation Loss. These categories quantify the overall Real-World (RW) SE as follows:

$$RW\ SE = (RUE) \cdot (RW\ Rank) \cdot Log_2[1 + (SNR - IL)] \quad (19)$$

Recall the focus of this paper has been on the DL, the RUE presented quantified the DL-RUE. This efficiency metric can also be applied to the UL (UL-RUE).

## VI. SE CONSIDERATIONS FOR 6G

We further discuss each contributing assumption in the overall practical deployment analysis. [24], [25]

### A. 5G Radio Resource Limitations

Recall the radio resource loss (= 1 – RUE) used in our analysis is 53%, which is significant. The individual components contributing to the RUE = 47% are TDD Slot Structure = 70%, Data Symbols = 78.6%, FEC Code Rate = 92.5%, Spectral Utilization = 98.2% and CP Overhead = 93.3%. The most significant impacting component is the TDD slot format. 5G radio resources loss can be reduced by allocating more downlink time slots, however this will reduce the uplink data rate – so this is not a suggested path forward. Here is a great application of Full Duplex (FD), which will allow the DL and UL to occur at the same time. It is expected that additional signal processing would be required to address any self-interference.[26], [27], [28] This approach will be able to improve that portion of the RUE from 70% to ideally 100%.

Another potential solution is to use more data symbols (less Pilot symbols) within the time slot, say 12 out of 14 total symbols. This would enable 85.7% of the downlink resources (instead of the assumed 78.6%). Recall we already considered 1 Pilot symbol + 11 data symbols, so this proposed solution would require a Pilot Symbol-Less communication link. We can also extend this to a CP-Less communication link. Here is a great application of an AI/ML based receiver.[19]

Just considering these comments we can improve the RUE from 47% to 78% (=85.7% * 0.92578 * 98.2% * 100% * 100%), and the radio resource loss (= 1 – RUE) from 53% to 22%. This is a significant overall gain in SE. We can carry this further, but requires removal of the control channel, and for this scenario, use all 14 symbols for data. A topic to be considered in defining an AI-Native communication system.[22], [23]

It is important to consider the efficiency of radio resources when delivering the expected services to the user. Based on the results presented in this paper, we have introduced a new System Metric to help define 6G: RUE. A proposed goal for 6G is provided in Table 3. With the assumed 5G radio resource configurations used in this paper, 5G yields an RUE of 47% with a corresponding radio resource loss of 53%. Using the proposed 6G goal of 1.5x's that of 5G, 6G can yield an RUE of 75% with a corresponding radio resource loss of 25%, basically doubling its contribution to SE.

TABLE III
Radio Resource Utilization Efficiency Goal for 6G.

| System Metric | 5G Goal | 6G Goal | Comments |
|---|---|---|---|
| RUE | N/A<br>(5G Baseline = 50%) | 1.5X's | 6G RUE = 75% |

### B. Channel Matrix Rank Deficiency

Since we have been focusing on SE, spatial multiplexing will allow us to maximize these gains by having a multiplier outside the "$\text{Log}_2(1+\text{SNR})$" function. Observing a full channel matrix rank is a very optimistic assumption. A solution would need to increase the rank of the channel matrix in locations that are rank deficient. This could be a great application of Reconfigurable Intelligent Surfaces (RIS), where instead of focusing on increasing SNR for coverage purposes (moving towards rank reduction).[5], [6] The proposal would be to use RIS to increase the channel rank (moving towards rank expansion) thus allowing for more spatial data streams to be supported resulting in higher spectral efficiency.

Another technique to support parallel data streams is to make use of Space Division Multiple Access (SDMA) using Massive MIMO to enable user-based Beam Forming, referred to as Multi-User MIMO (MU-MIMO). Here each user would receive parallel data streams via separate beams. This way low rank channels would not preclude the increase in SE since additional Degrees of Freedom (DoF), via the dimension of space, would be used for communication.

Table IV quantifies the overall SE loss when considering potential SE improvements in 5G radio resources (RUE) and channel matrix rank assumptions. We begin with the 5G baseline, RUE = 50% and the assumed 60/40 channel rank, which results in an aggregate SE loss of 65%. Next when considering a Pilot Symbol-Less receiver, reduced the aggregate SE loss to 62%. Lastly, the addition of Cyclic Prefix-Less and Full Duplex, resulted in a SE loss of 45%. Hence RUE improvements provide an overall SE gain of 20% (= 65% - 45%). If we assume by deploying RIS, the Channel matrix rank distribution can be improved to 20/80, then an overall SE gain of 35% (= 65% - 30%) may be achieved.

TABLE IV
Potential SU-MIMO SE Improvement Scenarios
(PS = Pilot Symbol, CP = Cycle Prefix, FD = Full Duplex)

| SE Components | 5G Baseline | PS-Less | PS-Less, CP-Less & FD | 5G Baseline | PS-Less | PS-Less, CP-Less & FD |
|---|---|---|---|---|---|---|
| RUE | 50% | 55% | 78% | 50% | 55% | 78% |
| Channel Rank Assumption | 60/40 | | | 20/80 | | |
| Aggregate SE Loss | 65% | 62% | 45% | 55% | 51% | 30% |

Our discussion has focused on a deployment where all 4 layers are allocated to a single user, referred to as SU-MIMO. Table V quantifies the expected SE loss for a MU-MIMO deployment configured for 2 users, each having 2 layers (totaling 4 layers). In this case, we assume 2 layers can be encountered 100% of the time. With these assumptions, the overall SE loss for 5G baseline case reduced from 65% (SU-MIMO) to 50% (MU-MIMO). Applying the potential RUE improvements resulted in an overall SE gain of 28% (=50% - 22%).

When considering these results, improving RUE provided the largest overall SE performance gain. Additionally, utilizing low rank deployment, not only reduces the dependency on observing a large channel matrix rank, but also improves the overall SE performance.

TABLE V
Potential MU-MIMO SE Improvement Scenarios
(PS = Pilot Symbol, CP = Cycle Prefix, FD = Full Duplex)

| SE Components | 5G Baseline | PS-Less | PS/CP-Less & FD |
|---|---|---|---|
| RUE | 50% | 55% | 78% |
| Channel Rank Assumption | | 100% | |
| Aggregate SE Loss | 50% | 45% | 22% |

### *C. Implementation Loss (IL)*

Potential IL improvements were provided in Table 2 and have shown a 2dB gain (going from 3dB to 1dB) can increase the SE by 29% for the Cell-Free scenario. Using linear interpolation, we can estimate the gain to be ~14% for every dB of IL improvement. Here advanced signal processing algorithms for both the transmitter and receiver are expected to provide the required gains, at the expense of additional complexity. Here is a great application of AI/ML based techniques for future communication systems based on AI-Native principles.[22], [23]

Throughput can also be increased using Higher Order Modulation (HOM), beyond 1024-QAM, to increase the bps/Hz achievable. Moreover, spectrally efficient modulation schemes that not only perform well at higher SNR, but also for low SNRs will prove to be beneficial.

The implication of a new waveform was not discussed directly but can help the current discussion since the % of useable 5G BW decreases as the transmission BW increases. In other words, the spectral utilization should increase to improve its contribution to the RUE.

## VII. Increasing Transmission Bandwidth (BW)

### *A. How does Bandwidth Impact Spectral Efficiency?*

So far, all the SE results have been presented for a BW = 100MHz. Next, we would like to discuss the impact of transmission bandwidth on SE.[20], [21] Recall the SE of a single cell SISO system with no interference present is:

$$C = \frac{R}{BW} = Log_2\left[1 + \frac{\sigma_s^2 \cdot |h|^2}{\sigma_n^2}\right] \quad (20)$$

This equation is rewritten to focus on throughput, R (bps). We consider the received signal power, $S_r$, which includes the path loss. The noise power can be expanded using noise power spectral density, $N_o$, and channel bandwidth, BW.

$$R = BW \cdot Log_2\left[1 + \frac{S_r}{N_o \cdot BW}\right] \quad (21)$$

Figure 7 shows the SISO throughput, R, as a function of the transmission BW for various distances from the cell center: 30, 60 and 200 meters. A constant received signal power, $S_r$, implies the SNR is not constant across all the BWs. In fact, the SNR degrades as the BW increases since more noise is now considered in the SNR denominator. This implies SE will degrade as BW increases. This is because the received signal power is kept constant as the BW increases.

These results help answer the question: when should the transmission BW increase ? For the case close to the cell center (30m), there is almost a linear gain with increasing BW. For a larger distance, we start to see compression behavior, meaning less than linear increase with BW. Lastly for the user away from the cell center (200m), we see very little increase in throughput when increasing the BW. These results lead to a conclusion to use higher BW when higher SNR is available to benefit from the almost linear increase in throughput.

An alternative way to consider the impact of BW on throughput is shown in Figure 8. This shows for users close to the cell center (30m), having a larger transmission BW will help increase throughput even though the SNR is decreasing. For users further away from the cell center (200m), having a larger transmission BW offers no significant gain. For a given location within the cell, increasing the BW decreases the SNR because $S_r$ is kept constant in the respective locations. This information will prove useful to MAC schedulers when maximizing DL throughput.

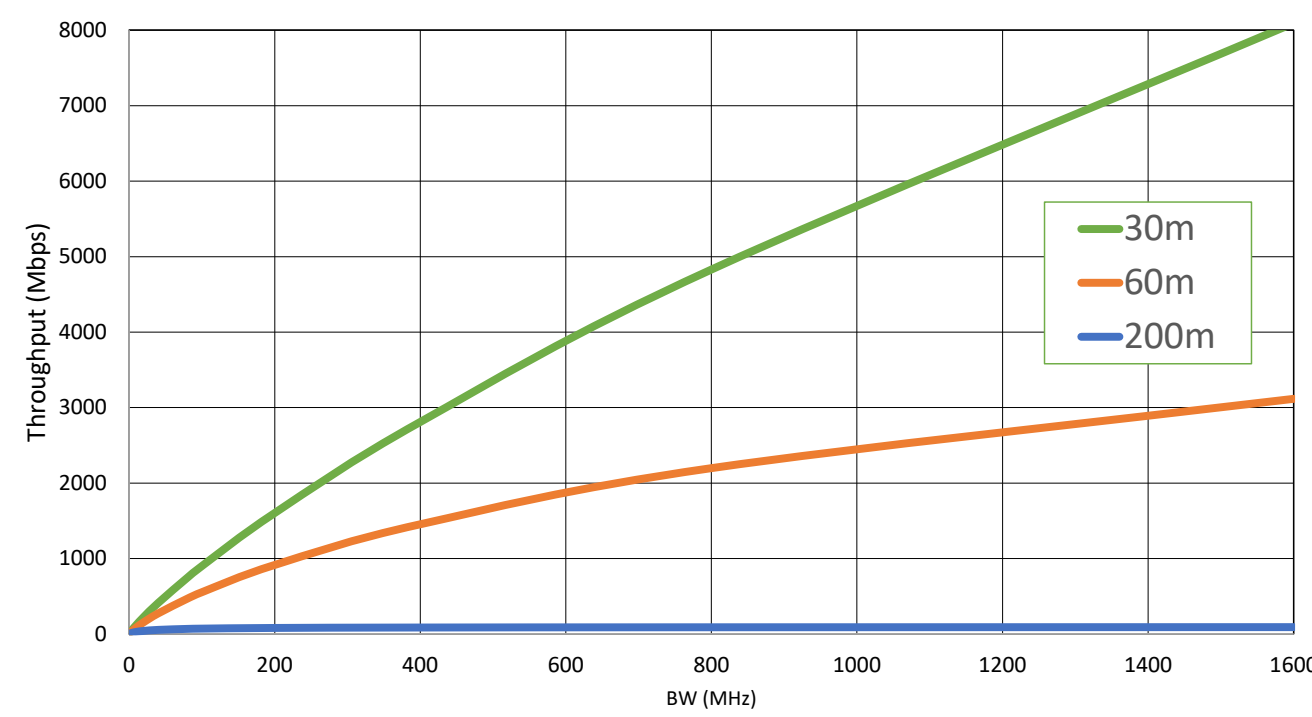

**Fig. 7.** Throughput performance for various BWs.

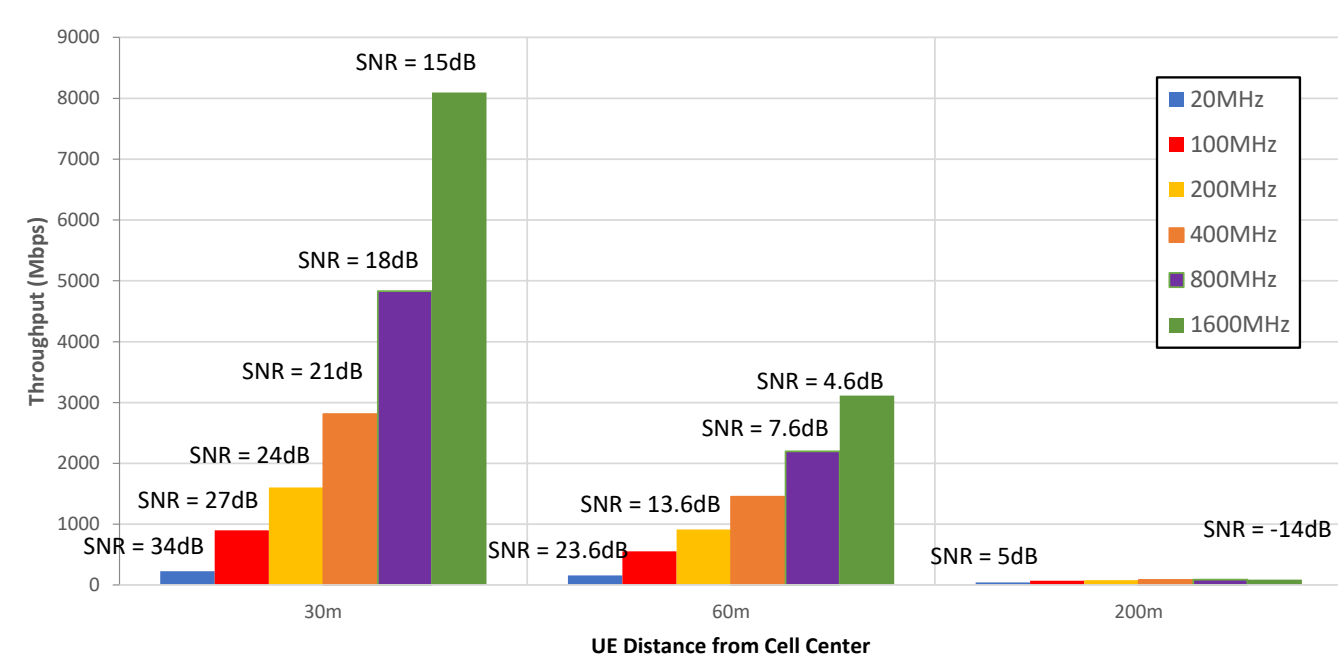

**Fig. 8.** SISO Throughput Comparison: BW and Locations within a cell.

A point worth mentioning, BW = 100MHz is the maximum allowed BW in the 3GPP defined FR1 frequency range. The higher BWs are allowed in the FR2 frequency range. Is it expected that different system design considerations (Tx antennas, Tx power, Path Loss, Rx antennas) will impact the respective link budgets of the various BWs. We have only considered the received signal power, $S_r$, in the analysis.

To summarize our findings: for a constant received signal power, $S_r$, lowering the BW will increase SNR. Whereas increasing the BW will lower SNR. Since SNR is proportional to SE, the same observations can be concluded when considering spectral efficiency – increasing the BW will lower SE. However, the throughput does increase with increasing BW (even with lower SNR), because of the multiplication of an increasing linear term proportional to the BW, as in (21).

The overall cell throughput can also increase by using Carrier Aggregation (CA). This technique allows simultaneous transmission across frequency carriers. The overall throughput increases by aggregating multiple carriers. However, CA will only increase the data rate and not change the SE. For example, consider 2x CA where each carrier will support 2 DL layers hence the data rate will double (equivalent to 4 DL layers), but twice the BW is needed to enable this increase. Here we considered each CA to support 2 layers, since it is easier to achieve than a single carrier 4 DL layer assumption.

Figure 9 illustrates the relationship between SE and cell throughput (Data Rate in Mbps) for all the BWs considered. Here we selected the maximum number of PRBs allowed for each transmission BW and assumed 256-QAM modulation. We have selected the TDD slot format as DDDSUUDDDD. The SE was calculated using the occupied BW, for example we used 98.28MHz for the BW = 100MHz label. The cell data rate was calculated using 27 out of the 40 time slots where 1 DL slot was left for SSB, SIB1, etc. communications.

First, consider an occupied BW of 100MHz, this curve shows as the number of layers increases, the spectral efficiency grows exponentially as well as the cell throughput. The points on the curve are represented as: 1 Layer SU-MIMO, 2 Layer SU-MIMO, 4 Layer SU-MIMO, 8 Layer MU-MIMO and 16 Layer MU-MIMO.

When comparing all the BWs (for a fixed user configuration), the spectral efficiency stayed approximately the same, however the data rate increases with the larger BW. To overcome channel matrix rank deficiency [17] & [18], a potential solution is to use CA. Here we begin with single carrier using 2 DL Layers and then increase the number of carriers to 2, 4 and finally 8. The CA curve "splinters" off the exponential curve to increase the data rate, however the spectral efficiency stays the same as indicated by the horizontal dashed line. These curves provide the best possible SE and Peak Cell data rate given the 5G RUE discussed, they do not include any implementation loss (IL). If we compare the 8 CA case to the 16 Layer MU-MIMO case, for BW=100MHz, we notice the cell throughputs are approximately equal; however, their spectral efficiencies are dramatically different.

Figure 9 provides an alternative viewpoint worth mentioning. For a constant SE value, increasing the BW will increase the throughput. Also, for a constant throughput, decreasing the BW will increase the cell Spectral Efficiency (if more layers are used). We have presented 4 choices to increase the cell throughput: Increase SINR, Increase the BW, Increase the MIMO layers and/or use CA.

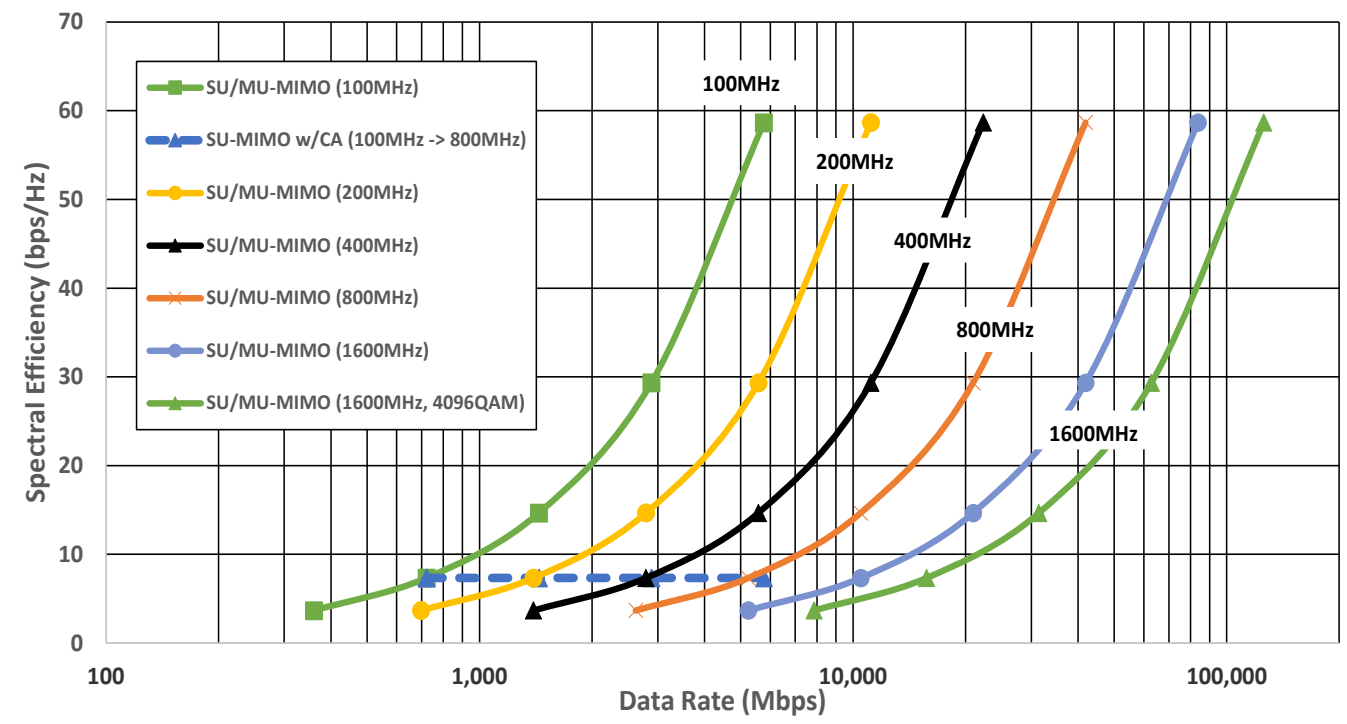

**Fig. 9.** SU/MU-MIMO Spectral Efficiency vs. Throughput for various transmission BWs.

### *B. Comparison to Commercial Network Measurements*

In this section we provide a summary of the channel rank measurement results obtained from 5G (& 4G) deployed networks.[17], [18]

For sake of comparison to measurements made in a 5G commercial network, we added 2 points on the 100MHz curve shown in Figure 10.[17] These data points are indicated by two red markers and labeled as Field Trial Results. The lower SE data point reflects the SU-MIMO results, where an average user data rate of 911Mbps measured on the DL, resulting in a SE of 911Mbps/98.28MHz = 9.27bps/Hz. The higher SE data point reflects the MU-MIMO results, where a total average of 2909Mbps was measured on the DL, resulting in a SE of 2909Mbps/98.28MHz = 29.6bps/Hz.

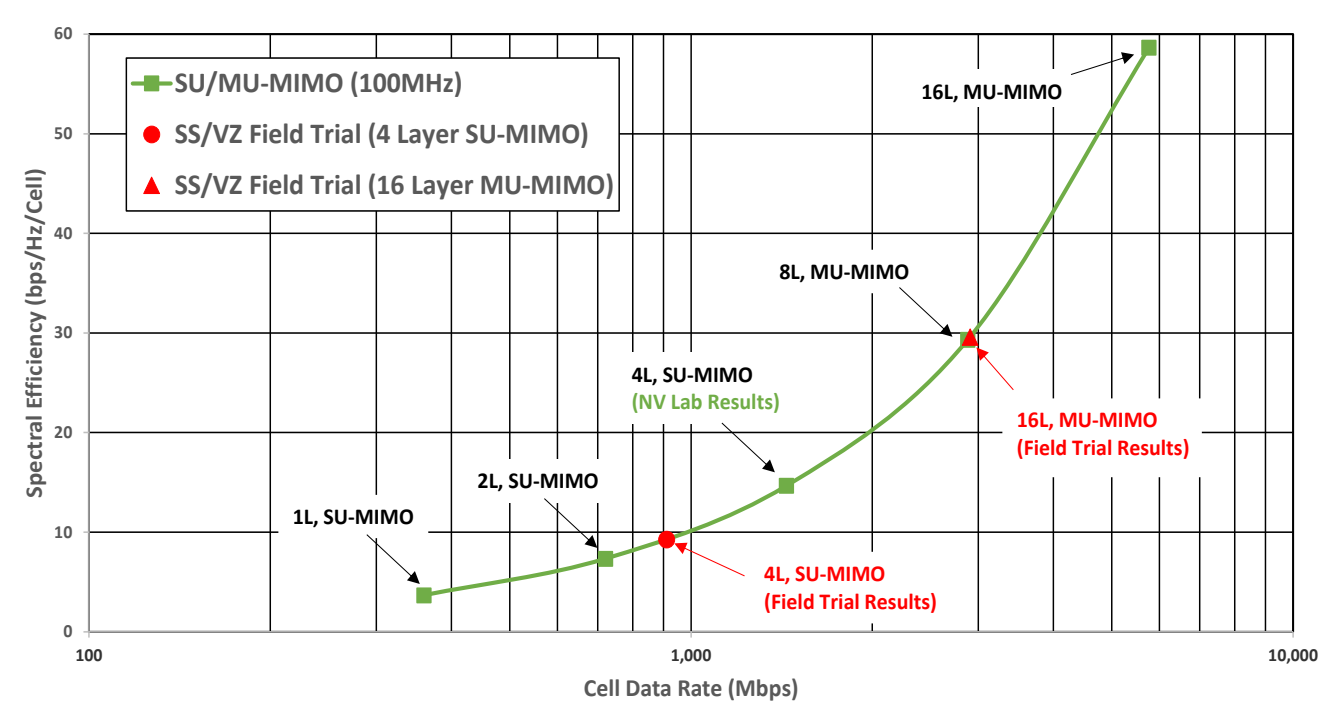

**Fig. 10.** SU/MU-MIMO Spectral Efficiency comparison to measurements.

The SU-MIMO results are lower than the max possible 4 Layer SU-MIMO expectations, this was because the average rank measured was approximately 3.4 and the maximum PRBs allocated in the commercial network was less than what was assumed in the remaining data points shown in Figure 10.

For MU-MIMO, 8 UEs with 2 Layers each were scheduled, which aggregated to 16 layers. The MU-MIMO results are lower than the max possible 16 Layer MU-MIMO expectations, this was because the total measured rank was approximately 13.4 and max PRBs allocated was less than what was assumed in Figure 10.

Next, other measurements were made on three commercial 4G & 5G networks on low and mid frequency bands. These results showed the channel matrix rank rarely supports the full 4 layers of a 4x4 MIMO link.[18] We averaged 6 data sets (provided in [18]), and the results revealed the probability of having a full rank was ~7% and the overall channel matrix rank was ~2.5.

## VIII. Centralized Network Architecture

In the previous sections we identified 3 categories that impact improvements in SE that need to be addressed in the definition of 6G. The proposed solutions require a network architecture that supports multi-cell signal processing and enables the adoption of AI/ML based techniques to realize the expected performance improvements. Multi-cell signal processing would best be served in a centralized location, called C-RAN (Centralized-RAN). Moreover, implementing the RAN components via a SW-defined network would be called Cloud-RAN. In fact, having a network architecture that can allocate network functions in a centralized fashion and a distributed fashion (such as to support edge compute) is a highly desirable capability in future deployments.[29], [30]

First, consider a typical cellular deployment, Figure 11 shows a RAN Network Architecture which consists of Central Processing Units (CPUs), Network Interface Cards (NICs) and fixed function HW accelerators. In this deployment, eleven cells are shown, each supported by a separate O-RU. We have chosen the left most five cells to be supported by a single O-DU and the remaining six right most cells to be supported by another O-DU. As a UE enters the cell (respective O-RU coverage area) the O-DU associated with that particular O-RU supports the UEs communication.

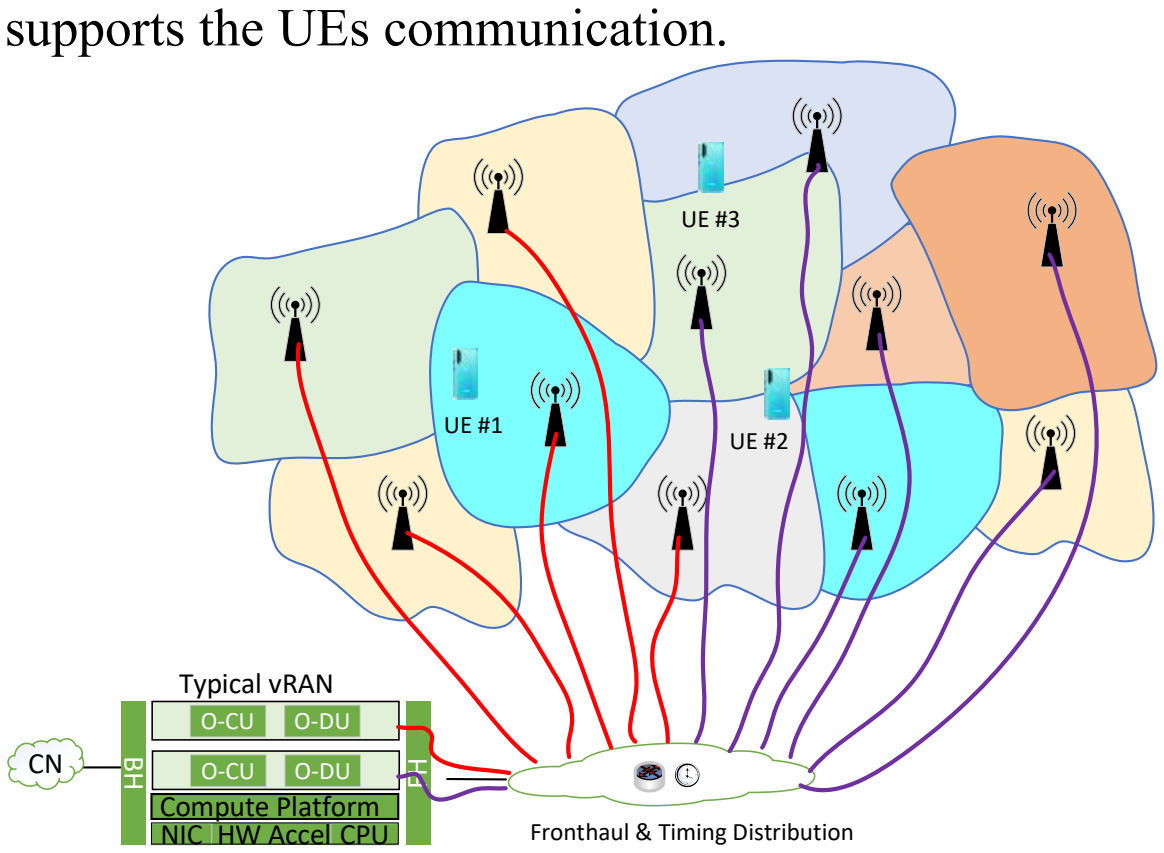

**Fig. 11.** Example RAN Network Architecture – Typical Cellular Deployment.

We introduced a paradigm shift causing system designers to rethink the evolution of typical cellular deployments. Generative AI is becoming more mainstream and as presented in this paper, plays a major role in considering improvements in SE. Hence having a Telecom network that can efficiently support the next generation RAN deployments as well as future Generative AI applications will be required. The AI-RAN Alliance [12] defined 3 scenarios: "AI-and-RAN" where the same server infrastructure is shared between AI and RAN workloads to increase utilization during low RAN traffic periods, "AI-on-RAN" where new AI based uses cases operate over 5G communications and "AI-for-RAN" where AI/ML algorithms are used to improve RAN performance, such as SE & EE. Hence the next generation Telecom networks need to efficiently support AI/ML workloads. To support a rapid rate of innovation, the solution needs to be SW-defined and performant. Besides the traditional cellular roadmap items the next generation Telecom network needs to handle new technologies such as RIS and Cell-Free deployments.

Figure 12 shows an example RAN Network Architecture configured for a Cell-Free deployment consisting of a GPU, Data Processing Unit (DPU) and CPU. To enable low latency access to applications (ex. AI/ML based) we have utilized Control & User Plane Separation (CUPS) by including the User Plane Function (UPF) in the RAN network. It is expected AI/ML will continue to be used throughout the RAN where efficiency, productivity and performance improvements can be achieved.

In the example Cell-Free deployment, 3 UEs are shown serviced by groups of O-RUs creating 3 Clusters. A Cluster is defined by a set (or group) of O-RUs used to communicate the desired signal to a specific UE. With this, Cluster #2 has six O-RUs associated with UE #2. The O-DU needs to send the respective DL signals to all six O-RUs. The O-DU also needs to receive the respective UL signals from the six O-RUs. These clusters are dynamic, meaning they follow the UE as its location moves. The # of O-RUs within a cluster can also change, hence the formation of these clusters is dynamic.

Having a O-DU/O-CU supporting many cells will simplify multi-cell signal processing as all the processing required to cover a specific area is essentially within the same O-DU server. Hence a high capacity (cell count & throughput), SW-defined and performant network solution is optimal for Cell-Free deployments. Here a single O-DU/O-CU would be able to cover a significant amount of area (via intra-server processing) or clusters. When a user crosses a coverage area serviced by two O-DU/O-CU, data needs to be exchanged between them (via inter-server processing). If the cell capacity of the O-DU/O-CU is small, then these inter-server data exchanges occur more often and increase xHaul traffic and possibly latency. This is especially true as the UEs travel with higher vehicle speeds and as more O-RUs are deployed within a coverage area.

This technology evolution will require a flexible transport network, specifically SW-defined xHaul (FH & BH). For example, when considering Cell-Free deployments, FH traffic will not only need to be sent to more O-RUs but should also be dynamic to "follow" the user clusters as they are created. This requires the FH traffic to handle point-to-multipoint (single O-DU to multiple O-RUs) and multipoint-to-point (multiple O-RUs to single O-DU) traffic which occurs when multiple O-RUs provide the desired signal within a cluster.

In a typical cellular deployment, a cell within a O-DU is connected to a single O-RU. The C/U-Plane packets are exchanged between a single source O-DU MAC address and a single destination O-RU MAC address.[30], [31] In our example, we have a O-DU support six O-RUs. Here each cell in the O-DU has a one-to-one mapping of MAC addresses.

In a Cell-Free deployment, a user serviced in a O-DU is connected to many O-RUs. The C/U-Plane packets are exchanged between a single source O-DU MAC address and many O-RU MAC address destinations. In our example we have shown a O-DU to support three clusters, where each cluster supports three to six O-RUs.

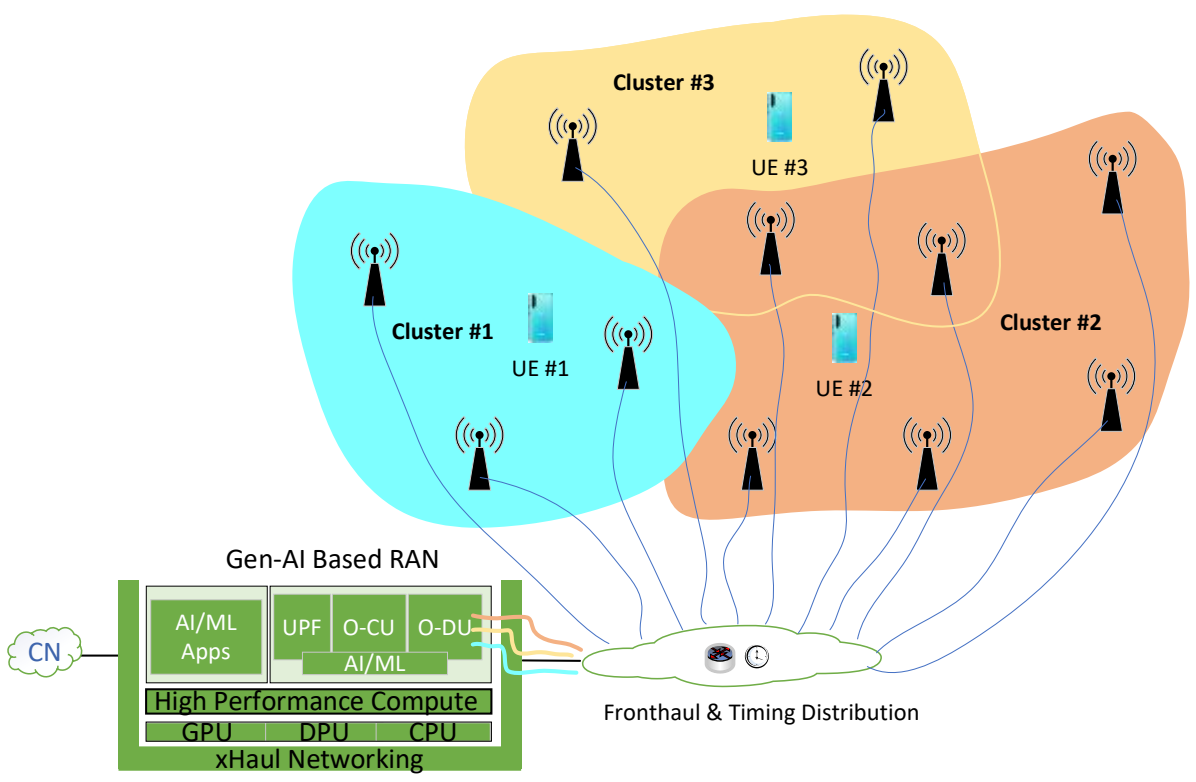

**Fig. 12.** Example RAN Network Architecture – Cell Free Deployment.

It is also worth mentioning the co-existence between traditional cellular deployments and the new deployments technologies will be needed. Thus, compounding the demand for a flexible transport network.

## IX. Conclusions

We provide a concise treatment of the items impacting SE by introducing 3 categories: 5G Radio Resources, Practical Limitations and Implementation Losses. The loss due to 5G radio resources is presented in the form of efficiency. We introduce the Radio Resource Utilization Efficiency (RUE) metric and show an example 5G system configuration yields an RUE of ~50%. As such, it is suggested that RUE be used in the definition of 6G to assist in meeting the next generation SE targets.

The loss due to Practical Limitations is considered as the channel matrix rank deficiency. We introduce the 60/40 notation to consider real world limitations. Results are compared to commercial measurements which show the likelihood of full rank channels is very low. When considering rank deficiency, the aggregated SE degradation increases from 50% to 65%. Including a 3dB IL, the aggregated SE degradation increases from 50% to 70% (for higher SNR values) and 80% (for lower SNR values), which are significant.

Theoretical DL spectral efficiency results are presented for two types of deployment scenarios: Typical Cellular and Cell-Free. When considering the impact of all 3 categories, The Cell-Free deployment results in approximately 3.83x's more SE at greater than 90% of the cell and 1.13x's more SE at greater than 10% of the cell, when compared to Typical Cellular. When only considering IL, an approximated SE gain of 14% is possible for every dB improvement in SNR.

We quantify the potential improvements in RUE from techniques such as Pilot Symbol-Less, CP-Less, Full Duplex, etc. The potential SE improvements from channel matrix rank deficiency is provided with the help of Reconfigurable Intelligent Surfaces. Also, with Advanced Digital Signal Processing via AL/ML techniques, IL improvements are quantified. When considering the proposed improvements for RUE and Practical Limitations, SE degradation with SU-MIMO are significantly decreased from 65% to 30%. Moreover, the use of lower rank communications with MU-MIMO reduced the SE degradation from 65% to 22%.

The transmission BW impact on spectral efficiency is presented to show an increase in data rate is possible. We show for constant receive power, $S_r$, the SNR decreases as BW increases. This in turn means the SE decreases as BW increases for constant receive power. Also, when considering various locations within a cell, the benefit of increasing the BW can be significant even when the SNR is at modest levels and the benefit can be insignificant when the SNR is low.

The performance gains in using multiple cells are worthy of a centralized deployment, such as C-RAN, to accommodate the necessary multi-cell signal processing. A SW-defined, HPC platform is proposed to support the co-existence of Typical Cellular and Cell-Free deployments as well as proliferate the adoption of Generative AI in the RAN.

We propose SE improvements, coupled with Cell-Free deployment and increased transmission BW, should be considered in the definition of 6G cellular technology. We expect SW-defined solutions to accelerate innovation.

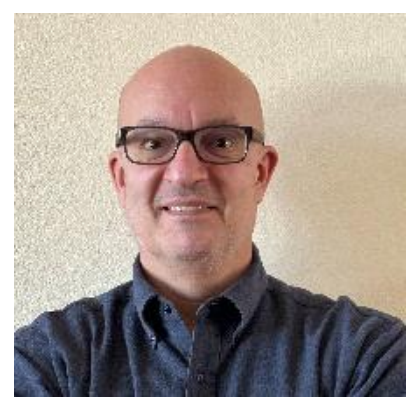

**Joseph Boccuzzi** (Member, IEEE) received the B.S. degree in Electrical Engineering from Polytechnic University of N.Y., M.S. degree in Electrical Engineering from Florida Atlantic University, FL. and Ph.D. degree in Electrical Engineering from New York University, Polytechnic school of engineering, N.Y.

He was employed by Motorola, AT&T Bell Labs, Cadence Design Systems, Morphics Technology (Purchased by Infineon Technologies), Broadcom, Mindspeed Technologies and Intel where he was the System Architect of Look-a-Side solution. He was an Adjunct Professor at NYU – Polytechnical School of Engineering where he created the "Wireless Digital Cellular Communications" graduate course.

He is with NVIDIA Corporation, Santa Clara, CA, responsible for End-to-End System Architecture for the Aerial AI-RAN 5G & 6G Wireless Platform (System Architect of In-Line solution). He has authored & co-authored several books: *Signal Processing for Wireless Communications* (McGraw Hill, 2009), *Femtocells: Design & Application* (McGraw Hill, 2010), and Chapter Author: *A Guide to the Wireless Engineering Body of Knowledge* (John Wiley & Sons, 2009), *Multiple Access Techniques for 5G Wireless Networks and Beyond* (Springer, 2018). His research interests include E2E Wireless Communications. Dr. Boccuzzi has filed more than 50 Patent applications.